\UseRawInputEncoding
\documentclass[a4paper,10pt,oneside]{article}
\usepackage[utf8]{inputenc}
\usepackage{icad2024,amsmath,epsfig,times,url,hyperref,amssymb}
\usepackage[T1]{fontenc}

\title{SOM\MakeLowercase{son} --- Sonification of Multidimensional Data in Kohonen Maps}

\twoauthors{Simon Linke} {Hamburg University of Applied Sciences \\ Ligeti Center\\ Hamburg, Germany  \\ {\tt simon.linke@haw-hamburg.de}}
{Tim Ziemer} {University of Hamburg \\ Institute of Systematic Musicology \\ Hamburg, Germany  \\ {\tt tim.ziemer@uni-hamburg.de}}

\begin{document}
\ninept
\maketitle
\begin{sloppy}
\begin{abstract}
Kohonen Maps, aka. Self-organizing maps (SOMs) are neural networks that visualize a high-dimensional feature space on a low-dimensional map. While SOMs are an excellent tool for data examination and exploration, they inherently cause a loss of detail. Visualizations of the underlying data do not integrate well and, therefore, fail to provide an overall picture. Consequently, we suggest SOMson, an interactive sonification of the underlying data, as a data augmentation technique. The sonification increases the amount of information provided simultaneously by the SOM. Instead of a user study, we present an interactive online example, so readers can explore SOMson themselves. Its strengths, weaknesses, and prospects are discussed. 
\end{abstract}

\section{Introduction}
\label{sec:intro}
Self-organizing maps \cite{kohonen}, also known as Kohonen maps, are artificial neural networks that represent a high-dimensional feature space on a low-dimensional map. This unsupervised learning technique can serve for data browsing, exploration and knowledge acquisition, pattern recognition, clustering, and data classification. SOMs are utilized in the fields of medicine \cite{sommedicine1, genetics}, biology \cite{sombio,somvet}, geology \cite{somgeo}, musicology \cite{hiphopsom,daga23}, sustainability \cite{somsus}, ethnology \cite{blassarchiv,kachin}, material science \cite{material,sommaterial} and many more.

In contrast to other artificial neural networks, SOMs are not a black box. Instead, they are explicitly designed to let users explore all network coefficients. Through analyzing the SOM, users gain an understanding of the training data. To date, this exploration and analysis is mainly based on visualization of single component planes or the somewhat condensed U-matrix (for a detailed explanation, see Section \ref{background}). While very useful and often intuitive, these visualizations cannot present the whole picture. As SOMs map a high-dimensional feature space to a two-dimensional grid, they do not simultaneously communicate all feature magnitudes of the underlying items. As a solution, we present SOMson, the sonification of a SOM based on a four-dimensional feature space. SOMson enhances SOMs by sonifying each node of the unit layer, allowing users to explore more aspects of the underlying data in an integrated fashion. Instead of evaluating the benefit of SOMson through a user study, we decided to provide a clear explanation of SOMs and SOMson and let the readers experience SOMson themselves in an interactive demonstrator.


In the remainder of the paper, we explain self-organizing maps, how they work, and how they are visualized. Then, we explain the psychoacoustic sonification, what data it sonifies, and how. After that, we present the SOMson user interface. Then, we give a guided tour through SOMson in an interactive online demonstrator, complemented by video examples. Finally, we discuss SOMson and provide a short conclusion.

\section{Self Organizing Map}
\label{background}
Self Organizing Maps (SOMs) are explained in detail by their inventor in \cite{kohonen}. We briefly summarize it here to give a better understanding of SOMson. SOMs are artificial neural networks with just one input and one output layer, the \emph{unit-layer}. 

In our demonstrator, we analyzed $15$ songs, generally referred to as items. From each song, we extracted the $4$ features \emph{PhaseSpace}, \emph{ChannelCorrelation}, \emph{PhaseSpaceHigh} and \emph{bpm} \cite{recordingstudiofeatures,daga23}. This way, each item is represented by a $4$-dimensional feature vector. Each song would have its unique location in a four-dimensional space, and their constellation could be observed in terms of proximity, like Euclidean distance. However, visualizing a four-dimensional space is not straightforward. Instead, a SOM is trained to represent the spatial constellation on a two-dimensional grid.

The $15$ items with their $4$-dimensional feature vectors are the input layer of the SOM. Each item from the input layer is connected to each node in the output layer, the so-called \emph{unit layer}. In our case, the unit layer is a quadratic grid with $16\times 16$ nodes. Every node holds a $4$-dimensional vector, a pointer into the four-dimensional feature space. We call the combination of a node and its pointer a \emph{unit}. Initially, the unit layer is randomized, i.e., each pointer points at a random location in the $4$-dimensional space. This unit layer will be trained iteratively, whereby the (now two-dimensional) representation of all $15$ items should preserve the original (high-dimensional) topography as closely as possible. This makes the SOM comparable to a projection of a high-dimensional space on a low-dimensional space or to a multidimensional scaling approach.

To train the unit layer, we identify the node whose pointer is most proximate to the location of item $1$. We can call this node the \emph{winning node} and refer to the combination of this node and its pointer as the \emph{Best Matching Unit} (BMU). Now, the item ``drags'' this node's pointer toward the location of the item. This means we modify the pointer to become a weighted mean value of the item's location and the original pointer. The weighting is called the \emph{learning coefficient}, and the pointer modification is called \emph{learning}. The pointers of all neighboring nodes are modified, too. The larger the distance between a node and the winning node, the lower the learning coefficient. The decrease in learning over distance is called a \emph{neighborhood function}, often modeled as a Gaussian function centered around the BMU.
This implies that the first item already alters the landscape of the complete unit layer from random pointers to a somewhat noisy Gaussian curve around item $1$.

The procedure is repeated with all items. This means that all $15$ items will affect the pointers of all nodes. You can imagine that items proximate to each other strengthen the pointing towards them, at least from the nodes nearby. In contrast, items from very different locations ``steal'' pointers from their nearby nodes. Pointers of nodes in between are torn back and forth, therefore pointing towards their middle. In machine learning terminology, this is referred to as \emph{competitive learning}.

In the first round, the learning coefficient is large, and the neighborhood function has a small decrease over distance. When all items have been used to train the SOM, the process is repeated. 
Every round, the learning coefficient is reduced, and the neighboring functions becomes narrower, making the learning related to an iterative process. This way, the unit layer (re-)organizes itself. Hence the term \emph{self organizing} map. Every round, the order of the items is chosen randomly to ensure the same influence of every item. Of course, the calculation of the BMU is renewed every round, too. Due to the reduction of learning coefficients and neighborhood functions, the unit layer converges to a nearly stable state. Thus, the training ends after multiple iterations ($2000$ in the given example) since every round would only slightly affect the BMUs, while the overall topology stays the same.

When the training is over, 
each node's pointer points at another location. Each item with which the unity layer has been trained has one BMU. Items that are similar in terms of many features will have proximate BMUs (or even the same BMU). They will cluster. Moreover, all pointers of the nearby nodes will point towards them. Items very different from this cluster but similar to each other (in terms of some feature magnitudes) may cluster somewhere else on the map. Again, all pointers of nodes around them will point towards this cluster. The nodes in between will point to the middle of these clusters and not to one of them. These are separation lines between two clusters. Depending on the data, items may not cluster but distribute over a subregion of the map. In this case, pointers may gradually point from item to item, like an interpolation. Such gradients do not exhibit clear separation lines.

Note that the unit layer is high-dimensional. Each node holds a pointer that is as high-dimensional as the input feature vector. However, the whole point of a SOM is the reduction of a high-dimensional feature space to a low-dimensional map that maintains the original topology. To date, this is achieved through some low-dimensional visualizations. Most importantly, the trained unit layer is visualized through the so-called \emph{U-matrix}. It is presented in Fig. \ref{pic:umatrix}. Instead of trying to visualize the $4$-dimensional pointer of each node, the U-matrix only shows the mean distance between a node's pointer and all neighboring nodes' pointers. When a unit and its neighbors point at the same $4$-dimensional location, it is plotted in black. The larger the mean distance between the node's pointer and the neighboring nodes' pointers, the lighter it is. This way, clusters appear as black islands, separated by white seas. The darker the island, the more similar the units. The lighter the sea, the larger the difference between the islands. On this map, each item from the training set is visualized by a colored dot on its respective BMU. Once trained, new items can be added to the SOM to analyze their relationship to the other items. 

\begin{figure}[ht]
\centerline{\includegraphics[width=83mm]{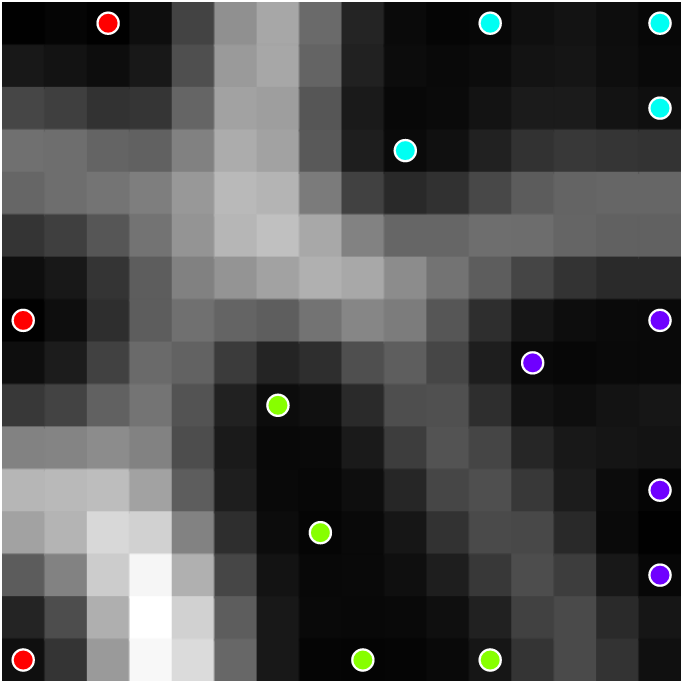}}
	\caption{{\it The U-matrix is the main output of a Self Organizing Map. Instead of visualizing the $4$-dimensional pointer at each of the $16\times 16$ nodes of the unit layer, the U-matrix indicates the mean distance between each node's pointer and the pointers of all its neighboring nodes. Nodes in the corners have $3$ neighbors, the other nodes along the fringe have $5$, and nodes in the middle have $8$ neighbors. The single items used to train the SOM (techno music) are shown as colored dots, whereby different colors represent different techno music styles.}}
	\label{pic:umatrix}
\end{figure}

In our example, we analyzed four different styles of techno music and plotted them in red, cyan, green, and blue. However, the training of the SOM is unsupervised, i.e., the algorithm is neither informed about, nor affected by our manual categorization. As you can see, the green dots cluster well. So do the blue dots. A gray separation line separates both clusters. The cyan dots cluster well, too. They are separated from the blue dots by a slightly lighter separation line, indicating that these clusters are a bit more different than the green and the blue one. The separation between the green and the cyan clusters is even lighter. The only white separation line can be found between the red item in the lower-left corer and the green cluster, indicating that this red item is more distinct from the green cluster than, for example, from the other red dot on the left-hand side of the map.

The U-matrix provides us with a lot of information about the relationships between the items. But there are four things we cannot see:
\begin{enumerate}
    \item Where in the feature space are our items allocated?
    \item In how far are items on the same island different from one another?
    \item In terms of which features are the islands different from one another?
    \item How similar or dissimilar are those islands that are no neighbors?
\end{enumerate}

The so-called \emph{component planes} are an approach to answering these questions. Component planes map the magnitude of a single feature at each node to color. Here, dark blue represents the lowest value, and yellow represents the highest value. The four component planes of our SOM are illustrated in Fig. \ref{pic:componentplanes}. The component planes reveal, for example, that the cyan items exhibit the highest \emph{Phase Space} value, the highest \emph{Channel Correlation}, the highest \emph{BPM} value, and a medium value of the \emph{Phase Space High} feature. So, overall, this cluster is somewhat extreme. In other words, the component planes allow us to answer question $1$. Moreover, we can see that cyan items mostly differ from each other in terms of \emph{Phase Space High}, which answers question $2$. When we compare the island of the blue items with the island of the cyan items, we see that both share a medium to high \emph{Phase Space High} and mostly a high \emph{Phase Space}. What distinguishes them the most is the \emph{Channel Correlation}, followed by the \emph{BPM}. This means we can answer question $3$. Last but not least, we can compare the islands that are no neighbors. For example, the uppermost red item and the island of blue items share a similar \emph{Channel Correlation}, while the \emph{Phase Space} value of the red item is much lower, and the \emph{BPM} value is just slightly lower. In contrast, the \emph{Phase Space High} value of the red item is a bit higher than that of the blue items. This answers question $4$.

\begin{figure}[ht]
\centerline{\includegraphics[width=83mm]{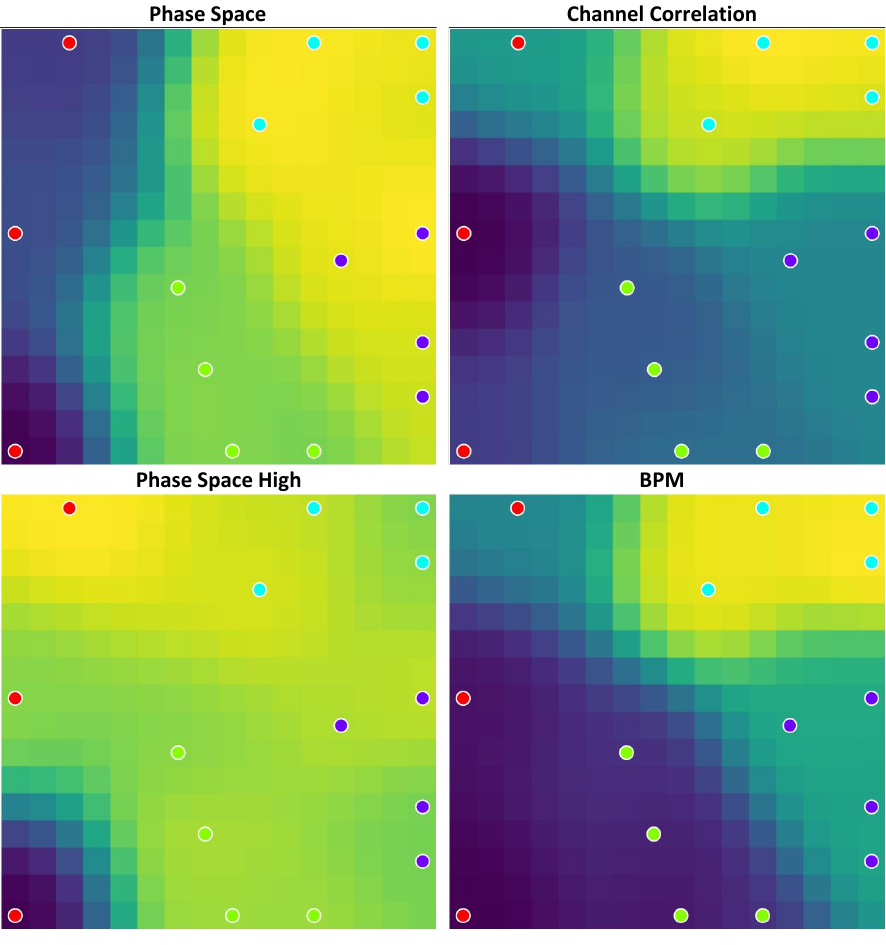}}
	\caption{{\it The component planes plot the magnitude of each one feature at each unit from dark blue (minimum) to yellow (maximum).}}
	\label{pic:componentplanes}
\end{figure}

Together, the U-matrix and the component planes present a lot of the information inherent in the SOM. The benefit is that once learned, they can be interpreted quite easily. Therefore, SOMs allow users to analyze data sets, identify clusters and distributions, e.g., to explore or presort new datasets, or browse through data based on similarity. Moreover, they allow researchers to study the explanatory value of single features and feature combinations, the intuitiveness of different distance measures, and much more.

The downside of these visualizations is that they do not provide the overall picture. In the U-matrix, most islands look the same, leaving the $4$ open questions listed above. Even though the component planes answer these questions, we can only analyze them one by one. Users either plot them on one screen and can only focus on one at a time, or they skip between graphics. All these visualizations do not integrate well. As a solution to this problem, we suggest SOMson, an interactive, multidimensional sonification of each unit's feature magnitudes. SOMson can be considered an augmentation of the SOM visualizations to increase the informativeness of Kohonen maps.

\section{SOM\MakeLowercase{son} Sonification}
\label{sec:pagestyle}
The simultaneous auditory display of multiple data dimensions or variables is often referred to as \emph{multidimensional}  or \emph{multivariate sonification} \cite{pampas,multivariate}. Multidimensional sonifications have already been proposed in 1980 in \cite{9dimensions}, and later, e.g., in \cite{asasonifi,davidneedle}. In the course of the Sonic Tilt Competition 2023
, many new two-dimensional sonifications with $2$ polarities each have been developed, like \cite{aeiou,voice,assist,bonk}. 
Important requirements for multidimensional sonification include 
\begin{enumerate}
    \item interpretability of dimensions
    \item continuity of dimensions
    \item linearity of dimensions
    \item a high resolution of dimensions and
    \item orthogonality between dimensions. \cite{ijis}
\end{enumerate}

The sound in our demonstrator is based on the psychoacoustic sonification as introduced for two dimensions in \cite{ziemerschicad}
and extended to three dimensions in \cite{icad2019}. Experiments with passive \cite{arxiv} and interactive users \cite{jmui2} have revealed that the psychoacoustic signal processing fulfills the above-mentioned requirements through a mapping of single dimensions to \emph{chroma}, \emph{roughness}, \emph{sharpness} and \emph{loudness fluctuation} of a Shepard tone. The orthogonality between these dimensions is also demonstrated in a YouTube playlist: \url{https://www.youtube.com/watch?v=7EeB7AGJnpQ&list=PLVv3BMS8IIXGo-SkwwD9rSUQKCPLy89kK}. Note that we treat roughness and loudness fluctuation as independent dimensions, while the cited literature uses them to represent two polarities of the same dimension.

Implementation of the psychoacoustic sonification is straight-forward: The source code can be found in \cite{sonictiltsourcecode} and has been implemented in the CURAT sonification game \cite{curat} and the Tiltification spirit level app \cite{tilt}. Experiments with the psychoacoustic sonification have shown that training is not necessary \cite{isontable} but helpful for the interpretation \cite{ziemerschicad} and interaction \emph{jmui}. 

What is sonified by the multidimensional sonification is the pointer at each node of the SOM. Here, each dimension of the pointer, i.e., each feature magnitude, is mapped to the magnitude of one dimension of the psychoacoustic sonification:
\begin{enumerate}
    \item \emph{PhaseSpace} $\rightsquigarrow$ Carrier Frequencies $\longrightarrow$ \emph{Chroma}
    \item \emph{ChannelCorrelation} $\rightsquigarrow$ Frequency Modulation Index $\longrightarrow$ \emph{Roughness}
    \item \emph{PhaseSpaceHigh} $\rightsquigarrow$ Peak Position of Amplitude Envelope $\longrightarrow$ \emph{Sharpness}
    \item \emph{bpm} $\longrightarrow$ Amplitude Modulation Frequency $\longrightarrow$ Speed of \emph{Loudness Fluctuation}
\end{enumerate}
where ``$\longrightarrow$'' indicates a linear mapping and ``$\rightsquigarrow$'' indicates a nonlinear mapping.

Each feature magnitude is normalized, resulting in a variable $x  \in [0,1]$, which is used to modulate the audible output signal $y(\omega, t)$, which is produced by nine sine-wave oscillators:

\begin{equation}
    y(\omega, t)=\sum_{i=0}^8 \hat{A}_i \ sin \left(2 \pi \ \omega_i \ t  \right)
\end{equation}

Each frequency $\omega_i$ is modulated by the \emph{PhaseSpace} $x_{\text{PS}}$ to produce a \emph{Chroma} as follows:

\begin{equation}
    \omega_i = 25 \cdot 2^{ \left(i+ \frac {4x_{\text{PS}}} {12} \right) }
\end{equation}

The Carrier Frequencies of the \emph{Chroma} are further modulated to add \emph{Roughness} representing the \emph{ChannelCorrelation} $x_{\text{CC}}$. 

\begin{equation}
    y(\omega, t)=\sum_{i=0}^8 \hat{A}_i \ sin \left(2 \pi \ \omega_i \ t + I(x_{\text{CC}}) \times sin(2\pi \times 30 t) \right)
\end{equation}
Where the modulation index $I$ is given by a mixture of a logarithmic and a linear mapping:
\begin{equation}
    I(x_{\text{CC}})=0.4\times5^{2.8 \ x_{\text{CC}}} + 0.6 \ x_{\text{CC}} \ 5^{2.8}
\end{equation}

\emph{PhaseSpaceHigh} $x_{\text{Ph}}$ is represented by modulating the Amplitude $\hat{A}_i$ of each oscillator according to its Frequency $\omega_i$ and thus, changing the \emph{Sharpness}:

\begin{equation}
    \hat{A}_i=\text{exp}\left( {-0.5 \left( 6.66 \ (\log_2(\omega_i)/9-(0.5+0.24 \ x_{\text{Ph}}))  \right)^2} \right)
\end{equation}

This Amplitude is further modulated by adding a signal $A_2$ with the same Amplitude and a frequency responding to the \emph{bpm}-feature $x_{\mathrm{b}}$:

\begin{equation}
    A_2=\hat{A}_i \times sin(2\pi\times 8 x_\mathrm{b})
\end{equation}

\section{SOM\MakeLowercase{son} Interface}

SOMson is designed as an interactive online demo that can be explored using a computer mouse. It is programmed using \emph{JavaScript}: \url{https://simon-linke.github.io/SOMson/} For sound synthesis, the \emph{p5.sound} library is used, which provides straightforward access to the \emph{Web Audio API} \cite{p5Sound}. (The link to the SOMson source code will be shared after the double-blind review process)

The SOMson user interface is fairly simple. A screenshot is presented in Fig. \ref{pic:interface}. It consists of the SOM visualization on the left and sonification parameters on the right. Six buttons allow switching between the U-matrix (\emph{Map}) and the component planes of the four components \emph{PhaseSpace}, \emph{ChannelCorrelation}, \emph{PhaseSpaceHigh} and \emph{bpm} as well as showing/hiding the training data.

\begin{figure}[ht]
\centerline{\includegraphics[width=83mm]{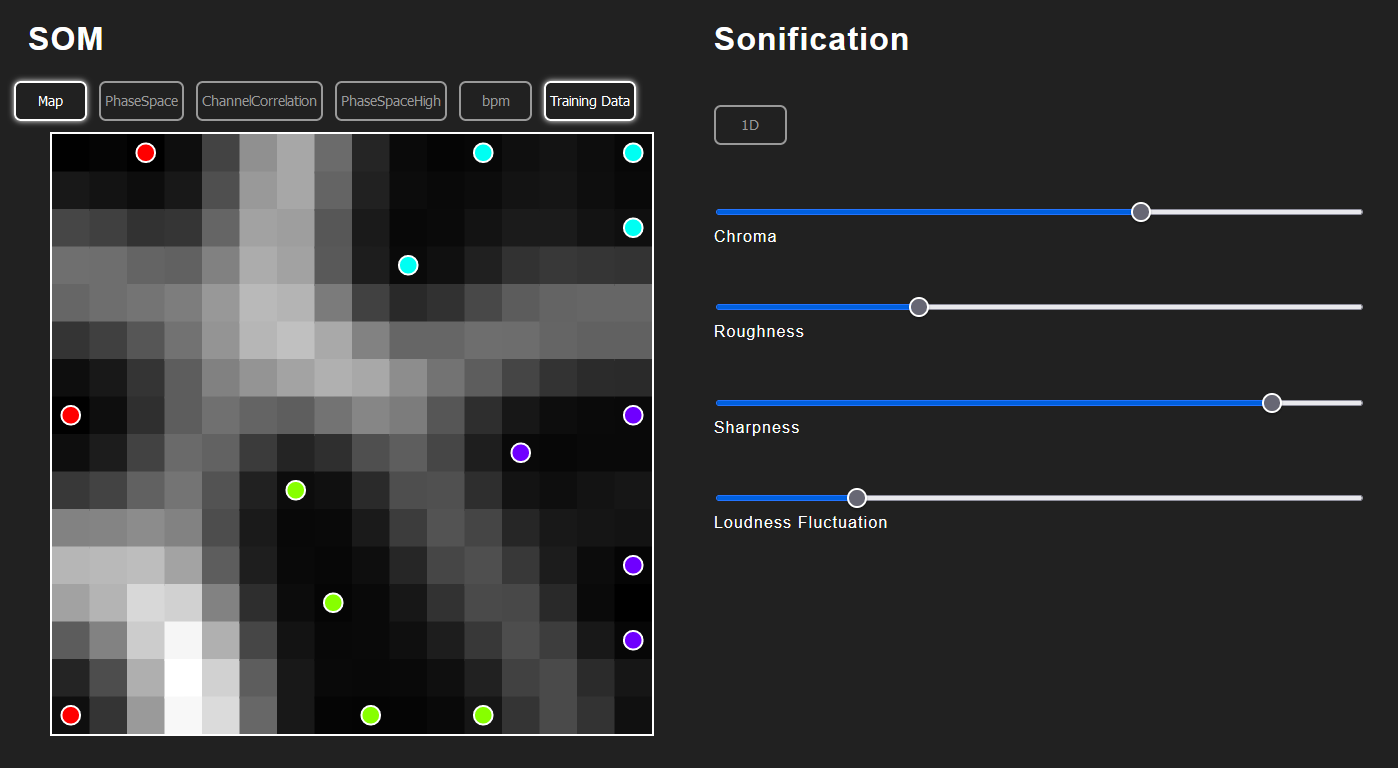}}
	\caption{{\it The SOMson interface with the visualizations on the left and the sonification parameters on the right. Buttons allow switching between U-matrix and the component planes, and showing/hiding the training data.}}
	\label{pic:interface}
\end{figure}

SOMson is controlled using a computer mouse. The sonification is interactive. Left-click a node on the map and hold the mouse button to hear the sonification of the unit. It represents the magnitudes of the unit, i.e., the four-dimensional pointer. Note that this information is not visible in either of the maps. Only the sonification provides this information. You can move from node to node to hear their difference. Only the node at which the cursor points will be sonified. This way, you can explore the map interactively. Release the mouse button to stop the sonification. This way, you can compare dedicated nodes or items.

The sonification sounds the same, no matter whether you load the U-matrix or any of the component planes. However, when one of the \emph{component planes} is loaded,  activating the \emph{1D} button will freeze all sliders except one. This way, the sonification is coherent with what you see.
This may also help in learning to distinguish the different sonification parameters. Using the mouse to move the sliders without clicking on the map will also change the selected parameter. Finally, while clicking on the map, the sliders on the right also move according to the magnitude of the selected units. This provided visual feedback may not significantly help explore the SOM's data, as it distracts attention from the map. Still, it provides helpful feedback when learning to distinguish the different audible parameters.

\section{SOM\MakeLowercase{son}: Guided Tour}
In this section, we guide you through SOMson. We recommend exploring our interactive SOMson project on \url{https://simon-linke.github.io/SOMson/simple/}. Alternatively, you can watch all single steps in our YouTube-playlist \url{https://www.youtube.com/watch?v=VqfHfaI_aVA&list=PLVv3BMS8IIXFoSn6p2svrIhNmPHupIW1c&index=1}.\\
\textbf{Step 1:} Compare items between two islands on the U-matrix: the red item in the upper-left corner and the green items. You may start comparing the red item with the uppermost green item. Click on the red one, then the green one, then the red one again. Repeat if you need more time. Concentrate on what has changed, by how much, and in what direction: pitch, roughness, sharpness, and loudness fluctuation. You may write down your observations. Then, explore the sound of all green items. What do they have in common, what is different, and by how much?

Differences between the islands that you can hear are:
\begin{enumerate}
    \item The red item has a much lower pitch than the green items
    \item The red item sounds sharper/brighter than the green items
    \item The loudness of the red item fluctuates much faster than the loudness of the green items
    \item The red item sounds subtly less rough than the green items
\end{enumerate}

Repeat your auditory inspection within all green items. Differences between items on a single island (the green ones) are:
\begin{enumerate}
    \item They have a very similar pitch
    \item They are audibly rough
    \item They all sound fairly sharp/bright (but the uppermost one is less sharp than the others)
    \item Their loudness fluctuates slowly, especially the item on the lower left
\end{enumerate}
\textbf{Step 2:} Stay at the U-matrix: The gray level indicates how similar neighboring fields are regarding the features with which the SOM has been trained. Now, explore the different fields on the map.

What you can hear is how similar the fields sound to their neighbors. Instead of summarizing all features to a single attribute (like the gray level in the visualization), the sonification indicates the magnitudes of all $4$ features with which the SOM has been trained. With some practice, you can hear out which and in how much the $4$ features have changed from one field to the next.\\
\textbf{Step 3:} In the U-matrix visualization, all clusters look alike: black islands mildly separated by gray lines or clearly separated by white lines. But sonification tells you more. Please explore how the reference island of green items (green island) sounds different from the island of purple and cyan items (purple and cyan island). Take your time and write down how chroma (pitch), sharpness (brightness), roughness, and loudness fluctuations differ. You can see that the green island is only mildly separated from the purple island, while the cyan island is clearly separated from the green island. Thanks to SOMson, you can also hear that
\begin{enumerate}
    \item the green island sounds more similar to the purple island than to the cyan island
    \item on the purple island, the intensity of all sound attributes has raised a bit
    \item on the cyan island, the chroma and sharpness have raised a bit, while roughness and loudness fluctuation have raised dramatically.
\end{enumerate}
Recall the mapping between feature and sound attributes. The sound indicates that songs on the cyan island are faster (bpm) and more mono (ChannelCorrelation) than the others.

Note that the respective video has some high-frequency artefacts that are not present in the interactive demo
.\\
\textbf{Step 4:} Pull all sliders to the left, and then gradually change the magnitudes of all $4$ parameters. You can hear that changing the magnitude of one parameter exclusively changes the intensity of one single sound attribute. It does not affect the others. This is one important requirement of multidimensional sonification: orthogonal dimensions must not interfere perceptually. Another requirement is that you can distinguish many levels of each parameter. This means the dimensions have a high resolution. The third requirement is that the dimensions are continuous: No interruption or \emph{jump} is heard when gradually moving the sliders. Last but not least, the relationship between the slider position and the perceived intensity of the respective sound attribute is linear: small motions always sound like small changes, no matter where the slider is located.\\
\textbf{Step 5:} Switch to the \emph{PhaseSpace} component plane, check the \emph{1D} box, and explore the first dimension. Before you do so, you may move all sliders towards the left for a more pleasant sound.

When browsing through the map, you hear the pitch change from low (dark blue) to high (yellow). You can simultaneously see and hear the magnitude of this dimension: some fields are the same, some are very similar, and some gradual changes occur.

You can make very many observations. For example,
\begin{enumerate}
    \item some purple items are more closely related to cyan items than to the remaining purple items
    \item even though far apart on the map, the red items are fairly similar to each other
    \item the red islands have a much lower pitch than all other fields.
\end{enumerate}
\textbf{Step 6:} Switch to the \emph{ChannelCorrelation} component plane, check the \emph{1D} box, and explore the second dimension.

When browsing through the map, you hear the clean sound in the dark blue region. Subtle differences between dark blue fields of different shades are audible. When moving from blue via turquoise to yellow, the roughness increases. Fields with similar colors also exhibit a similar degree of roughness.\\
\textbf{Step 7:} Browse through the \emph{PhaseSpace} component plane in \emph{1D} mode. Most fields on the map have a large value, yielding similar colors between lime green and yellow and a bright timbre between quite sharp and really shrill.  Visually and auditory, the largest contrast can be found in the lower-left corner. Here, the color and sharpness do not gradually fade but exhibit obvious steps. From lime green to dark blue and from quite sharp to dull. This effect stays audible, no matter what chroma, roughness level, or loudness fluctuation frequency you choose.\\
\textbf{Step 8:} Explore the \emph{bpm} component plane. The feature magnitude of bpm is mapped to the speed of loudness fluctuation. In the dark blue region, the loudness fluctuates very slowly. The fluctuation is getting faster over blue, turquoise, and lime green to yellow. Through this mapping, you can easily compare different fields on the map. Even though some turquoise fields look the same, you can hear which one fluctuates faster.\\
\textbf{Step 9}: Hear where seeing tricks you. Go, e.g., to the \emph{PhaseSpace} component plane and click on two spatially separated fields that you think look the same. You will realize that they often do not sound the same. You can hear which one has a higher magnitude, i.e., a higher pitch.

Such comparisons are of particular difficulty in vision, as human color and lightness vision is affected not only by the focused color but also by the relation to its neighboring colors and lightness levels, as exemplified in Figs. \ref{pic:colorillusion} and \ref{pic:brightnessillusion} in color (comparable to the component planes) and in grayscale (comparable to the U-matrix). This is one of the reasons why interactive sonification has been proposed as a complement for lightness, color, and contrast enhancement of visualizations \cite{niklasjmuicolor}.
\begin{figure}[ht]
\centerline{\includegraphics[width=83mm]{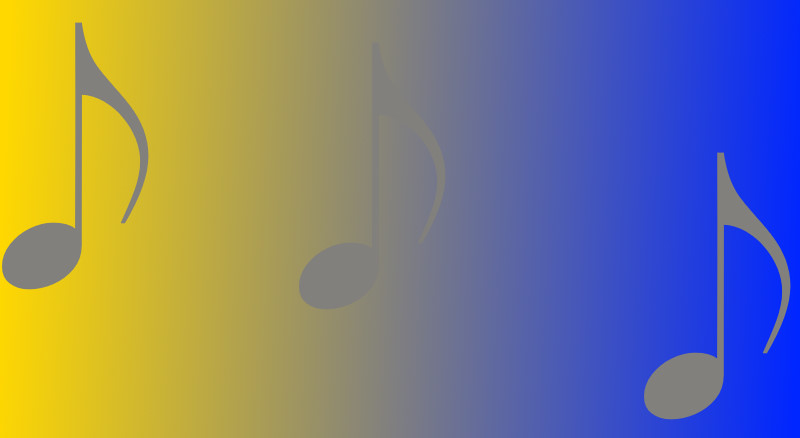}}
	\caption{{\it Bezold effect: Colors and shades may appear different depending on their surrounding colors. Here, all three notes have the same (single) color, even though the one on the left may appear darker, and the one in the middle may seem to have a color gradient.}}
	\label{pic:colorillusion}
\end{figure}
\begin{figure}[ht]
\centerline{\includegraphics[width=83mm]{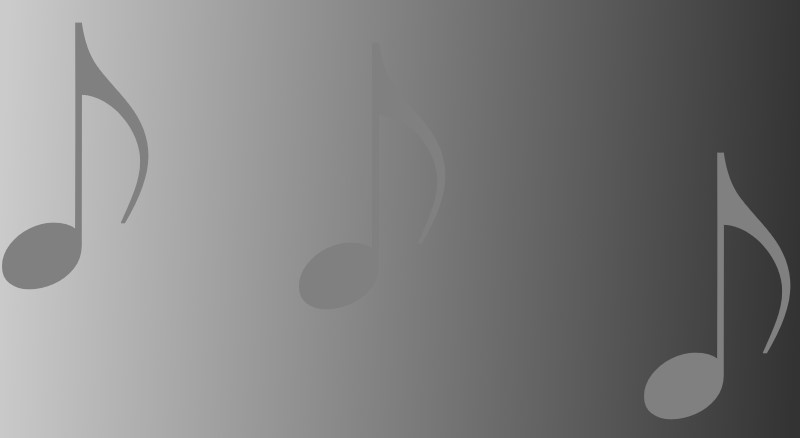}}
	\caption{{\it The Bezold effect also holds for lightness: All notes have the exact same (single) color and lightness level, even though the one on the left may appear darker and the one in the middle appears as if it had a lightness gradient.}}
	\label{pic:brightnessillusion}
\end{figure}

\textbf{Step 10:} Now that you have experience using and interpreting SOMson, you should explore and understand the underlying, invisible data of the SOM. For example, click on the red item on top to hear its feature magnitudes. If you have a feeling for the sounds already, you may realize that its pitch is on the lower side, its sharpness is very high, and the other attributes are somewhere in the middle.

Next, you should compare it to the red item in the middle. In what respects is it different?
\begin{itemize}
    \item It has a  (slightly) higher pitch
    \item It sounds (slightly) rougher
    \item It sounds less sharp
    \item Its loudness fluctuation is slower
\end{itemize}
Now, compare the red item in the middle with the red item at the bottom. How is the one at the bottom different?
\begin{itemize}
    \item It has a much lower pitch
    \item It sounds rougher
    \item It sounds way less sharp
    \item Its loudness does not fluctuate (magnitude of $0$).
\end{itemize}
When you listen with care, you will be able to interpret and compare the feature magnitudes of all items and nodes.


\section{Discussion}
Even though one of the authors had no previous experience with psychoacoustic sonification, both of us could work with it right away. It may take a while to a) learn on what sound aspects to concentrate on, b) get a feeling for the absolute magnitude of single features, and c) manage to integrate all sound attributes to get an overall picture of the feature magnitudes. But we could instantly hear a) which items were similar and which were not, b) where islands and seas were, and c) whether items differed in pitch, the speed of loudness fluctuation, or were alike. So, some of the information added to the SOM via sonification integrates seamlessly into the workflow, so gathering the necessary experience is only a matter of practice time. The sliders provided for each sonification parameter can further help reduce a user's adaptation time by providing visual feedback while exploring the map and allowing users to manipulate each parameter individually.

Of course, there are attempts to visualize multiple features of a SOM in a single representation. E.g., \cite{ponmalai2019} proposes to map the magnitude of the features to specific color channels of the RGB colorspace. An interactive example of this approach is shown here: \url{https://musicai.uni-hamburg.de/en/how-does-a-kohonen-map-work/}. Nevertheless, even this straightforward approach must be learned, as some experience is needed to recognize the single color channel from any given color. Further, it limits the dimensions of the feature space to a maximum of three.

Note that when using the auditory instead of visual parameters, a four-dimensional sonification is not a limit at all. For example, amplitude-based panning can be utilized to localize the Shepard tone at different azimuth angles, referred to as \emph{auditory event angle} in psychoacoustic terms \cite{ziemersonificationterminology}. This would yield a five-dimensional sonification. Another dimension that has already been suggested \cite{icad2019} and evaluated \cite{jmui2} is the auditory \emph{fullness} that can be implemented by reshaping the spectral envelope of the Shepard tone. Fullness is largely independent of the other five dimensions and fulfills the above-mentioned requirements for multidimensional sonification.

We exemplarily implemented a $7$-dimensional sonification available on \url{https://simon-linke.github.io/SOMson/extended/}. Here, a second auditory stream based on a noise generator is added. The first dimension of the noise is its color. It varies from brown over pink, white, and blue to purple. This mostly affects its brightness. Panning is utilized to implement the second dimension of the noise, which affects the auditory event angle. In the terminology of auditory scene analysis, the Shepard tone and the noise are segregated auditory streams \cite{ziemersonificationterminology}. This means that not only the sound generator but also the noise generator are distinct sound sources. More importantly, it means that the Shepard tone and the noise tend to be perceived as individual sound sources. The strategy of mapping multivariate data to attributes of various auditory streams has already been proposed in \cite{asasonifi}. The benefit of this segregation is that we can add more dimensions to the sonification without producing perceptual interference. A disadvantage is that you cannot accurately interpret several auditory streams' attributes simultaneously. To hear details, you have to concentrate on one stream and, if required, switch attention to the other. Sometimes, you can easily control your focus of attention. However, especially when drastic changes occur, the sound itself may capture your attention. Adding even more dimension is undoubtedly possible. But at some point, these pieces of information do not integrate well, meaning that the benefit of SOMson gets lost, which is presenting an integrated overview of all features of the SOM and its items. Furthermore, interpreting more dimensions requires better listening skills, more cognitive resources, and, certainly, more training.

In the $7$-dimensional demo, we also added the features to a \emph{Modulation Matrix}: 1.) Some mappings between data features and audio parameters are particularly intuitive, such as mapping \emph{bpm} to the speed of loudness fluctuations, where faster fluctuations mean faster music. We, therefore, allow mapping between features and sound parameters to be reassigned. 2.) Sometimes, it makes sense to invert the polarity, e.g., mapping the magnitude of a (hypothetical) \emph{darkness} feature to auditory sharpness/brightness from bright to dull instead of dull to bright. We, therefore, allow each mapping polarity to be inverted. 3.) Sometimes, listening to $7$ parameters at once is overwhelming, or the presence of one sound attribute distracts from the others. We, therefore, allow muting selected features.

So far, we have implemented SOMson using the p5.sound library \cite{p5Sound} and SOM data in JavaScript Object Notation (JSON), as Web Audio is widely accessible \cite{hansicad}. As using Python for data sonification is becoming more and more popular these days \cite{davidbuch,pyison,sonipy,strauss}, we are considering implementing SOMson as a Python package, too.

The SOMson source code is available on gitHub: \url{https://github.com/Simon-Linke/SOMson}

\section{Conclusion}
In this paper, we introduce SOMson, a sonification of self-organizing maps. As neither the U-matrix nor the single component planes provide all information about the underlying feature magnitudes, we provide them by means of interactive sonification. Based on a four-dimensional sonification, we guide readers through SOMson. With this interactive demonstration, readers can experience its benefits rather than imagining it based on demo videos or experiment results. We have developed SOMson for up to $7$ dimensions, but with increasing dimensionality, the interpretability requires more listening expertise, becomes more cognitively demanding, and not all aspects of the sound integrate well.

\section{ACKNOWLEDGMENT}
\label{sec:ack}
This research was funded under the Program "Innovative Hochschule" (innovative university) by the Federal Ministry of Education and Research (BMBF) of Germany (Grant No.13IHS232C) and the City of Hamburg. We thank Michael Blaß for his amazing Apollon package that includes a Self Organizing Map, which we used for generating our map.
\bibliographystyle{IEEEtran}
\bibliography{article}
\end{sloppy}
\end{document}